\documentclass[wcp]{jmlr}

\usepackage{longtable}

\usepackage{booktabs}

\usepackage[T1]{fontenc}    
\usepackage{hyperref}       
\usepackage{url}            
\usepackage[algo2e]{algorithm2e}

\usepackage{booktabs}       
\usepackage{amsfonts}       
\usepackage{amsmath} 
\usepackage{nicefrac}       
\usepackage{graphicx}
\usepackage{microtype}      
\usepackage{booktabs}
\usepackage{multirow}
\usepackage{graphicx}
\usepackage{comment}
\usepackage{algorithm}
\makeatletter
\let\Ginclude@graphics\@org@Ginclude@graphics 
\makeatother

\jmlrvolume{157}
\jmlryear{2021}
\jmlrworkshop{ACML 2021}

 \title[Beyond $L_p$ clipping]{Beyond $L_p$ clipping: Equalization-based Psychoacoustic Attacks against ASRs}

\author{\Name{Hadi Abdullah} \Email{hadi10102@ufl.edu }\\
  \Name{Muhammad Sajidur Rahman} \Email{rahmanm@ufl.edu }\\
  \Name{Christian Peeters} \Email{cpeeters@ufl.edu }\\
  \Name{Cassidy Gibson} \Email{c.gibson@ufl.edu }\\
  \Name{Washington Garcia} \Email{w.garcia@ufl.edu }\\
  \Name{Vincent Bindschaedler} \Email{vbindsch@cise.ufl.edu }\\
  \Name{Thomas Shrimpton} \Email{teshrim@ufl.edu }\\
  \Name{Patrick Traynor} \Email{traynor@ufl.edu }\\
  \addr University of Florida}
\begin{document}

\maketitle
\date{\vspace{-5ex}}
\vspace{-2em}
\begin{abstract}
Automatic Speech Recognition (ASR) systems convert speech into text and can be placed into two broad categories: traditional and fully end-to-end. Both types have been shown to be vulnerable to adversarial audio examples that sound benign to the human ear but force the ASR to produce malicious transcriptions. Of these attacks, only the ``psychoacoustic'' attacks can create examples with relatively imperceptible perturbations, as they leverage the knowledge of the human auditory system. Unfortunately, existing psychoacoustic attacks can only be applied against traditional models, and are obsolete against the newer, fully end-to-end ASRs. In this paper, we propose an equalization-based psychoacoustic attack that can exploit both traditional and fully end-to-end ASRs. We successfully demonstrate our attack against real-world ASRs that include DeepSpeech and Wav2Letter. Moreover, we employ a user study to verify that our method creates low audible distortion. Specifically, 80 of the 100 participants voted in favor of~\textit{all} our attack audio samples as less noisier than the existing state-of-the-art attack. Through this, we demonstrate both types of existing ASR pipelines can be exploited with minimum degradation to attack audio quality.
\end{abstract}
\section{Introduction}
 
ASRs have become increasingly popular as they can enable
users to seamlessly interface with their devices. This has greatly increased
accessibility by improving authentication and communication with users and their electronic devices. However, ASRs are vulnerable to adversarial audio
samples~\cite{schonherr2018adversarial,qin2019imperceptible,carlini2016hidden,carlini2018audio,abdoli2019universal,kreuk2018fooling,cisse2017houdini}. These are inputs that sound benign to human listeners but force ASRs to produce malicious transcriptions. This allows attackers to force home assistants to execute arbitrary
commands~\cite{biometrics,yuan2018commandersong} or evade surveillance
systems~\cite{abdullah2019kenensville}. ASRs can be categorized into two broad
types: traditional~\cite{arxiv-DS1,kaldi_hmm,sphinx} and
fully
end-to-end~\cite{sainath2015learning,jaitly2011learning,palaz2013estimating,tuske2014acoustic,hoshen2015speech,zeghidour2018end,zeghidour2018fully,fu2018end}. While traditional ASRs use signal processing algorithms for feature extraction (e.g., STFT), the newer end-to-end models use additional trainable layers that~\textit{learn} to extract the correct features. 
This divergence from a signal processing based feature extraction has made most of the existing attacks obsolete~\cite{qin2019imperceptible,schonherr2018adversarial} to the newer end-to-end ASRs. Unfortunately, the remaining attacks~\cite{carlini2018audio,abdullahs2020sok} that do work against the newer models produce low quality adversarial audio samples. Even though, past researchers have used the $L_p$ clipping to control the attack audio quality~\cite{abdullahs2020sok}, we show in this paper that such strategies are inherently flawed. They fail to consider psychoacoustics (i.e., the functioning of the human ear), resulting in noisy audio. 

We overcome these limitations and present a single unified attack that not only works against both ASR types (traditional and end-to-end) but also creates high quality adversarial audio. Our equalization-based attack uses the masking thresholds generated via the psychoacoustic model to control the quality of the perturbations. Constraining via the masking thresholds ensures that resulting perturbations will be largely imperceptible to the human ear.
We evaluated our attack against both traditional and fully end-to-end ASRs. Specifically, we attacked DeepSpeech (traditional) and Wav2letter (fully end-to-end) and achieved success rates of 100\% and 76\%, respectively. Upon further investigation, we observed that these success rates are comparable to existing optimization attacks~\cite{carlini2018audio} on such architectures. We then ran a user
study to determine whether our attack produces less low audible distortion than
the existing state-of-the-art attack due to  Carlini and Wagner (CW)~\cite{carlini2018audio}, which can also exploit end-to-end models. However, in contrast to our work, their attack does not capitalize on psychoacoustics to minimize audible distortion. Our user study
showed that 80 of the 100 participants found \textit{all} of our attack
audio samples to be less noisy when compared to those of the CW attack. Our work
demonstrates that, while improvements to ASRs architectures have made many
existing attacks obsolete, they remain vulnerable to attacks.

\section{Related Work}

The process of generating adversarial audio samples has generally followed one
of two approaches. The first is to model the ASR as an optimization function. Since
the ASR learns an optimal mapping between audio and intended output, attacking
the optimization directly is a viable approach when given knowledge of the target ASR
internals~\cite{bispham2018taxonomy}. With such knowledge, an adversary can
directly optimize an adversarial objective function that maximizes the error of
the ASR model’s objective function~\cite{carlini2016hidden}. This is practical in a
variety of settings, including voice assistants~\cite{carlini2016hidden,
yuan2018commandersong}, speech-to-text systems~\cite{cisse2017houdini,
schonherr2018adversarial, carlini2018audio}, music content
analysis~\cite{kereliuk2015deep}, and speaker-verification
systems~\cite{kreuk2018fooling}. The second is the signal
processing approach, which assumes that certain physical properties of audio
are modeled by the ASR. Often, the model of these physical properties is
imperfect, or does not faithfully re-create signals that are likely to be 
encountered in the
wild~\cite{vaidya2015cocaine,zhang2017dolphinattack,du2019sirenattack}. An
attacker can also leverage properties of the audio pipeline (e.g., FFT windows,
MFCC parameters) to generate adversarial audio directly for a desired
psychoacoustic effect~\cite{biometrics}.

Attacks that leverage knowledge of human psychoacoustics may produce higher quality
adversarial audio. These ``psychoacoustic'' attacks use the information about the human ear to craft samples with low audible distortion. Thus, to a human, the
resulting adversarial audio samples sound similar to the original
audio. These attacks a) assume knowledge of the model
internals, and b) leverage the underlying psychoacoustic properties that
control the modeling of audio. The knowledge required for such attacks
can vary. Researchers have proposed certain white-box
approaches~\cite{qin2019imperceptible,schonherr2018adversarial} which
assume that the ASR is using a particular signal processing algorithm (e.g., STFT) is being
used for feature extraction. Our work relaxes this
requirement and simply assumes the victim model is a fully end-to-end ASR model,
which need not contain any such signal processing algorithm for feature
extraction.

\section{Background}
\label{sec:background}

\subsection{Short-Time Fourier Transform}\label{stft}

The Short-Time Fourier Transform (STFT) provides the frequency information of a given signal over time (i.e., a spectrogram). This is accomplished via the sum of a sequence of Fourier transforms of overlapping windowed blocks of the time-domain signal. In this paper, we denote the application of an STFT over an audio sample $\vec{x}$ as: $ S(\vec{x}) = \text{STFT}_{m,n}\{\vec{x}\} $ . The STFT output is a $m \times n$ matrix of complex values and $m$ and $n$ are the time and frequency indices respectively.

For many applications, it is convenient to modify the STFT representation of the signal directly (as opposed to its time-domain representation). These include noise cancellation~\cite{boll1979suppression}, source separation~\cite{virtanen2007monaural}
, time-scale and pitch-scale modifications~\cite{laroche1999improved}. Typically, only the squared magnitude of an STFT is modified,
and the phase is either left unaltered or discarded~\cite{perraudin2013fast}. It
is important to note that, due to the overlapping frames of the STFT, modifications of the magnitude prevent a perfect STFT inversion (i.e., STFT is non-invertible). 

%
The most fundamental and widely used~\cite{shen2018natural,wang2017tacotron} technique for spectrogram inversion is the Griffin-Lim algorithm~\cite{griffin1984signal}. This is an iterative algorithm that \textit{approximates} the time-domain representation of the spectrogram. This technique is widely regarded as the most effective method of reconstructing an audio signal from its modified spectrogram. We denote this operation as: $ \vec{x}' = \text{Griffin-Lim}(S, k) $ . Here, $\vec{x}'$ is the approximated time-domain signal for spectrogram $S(\vec{x})$, and $k$ is a parameter of the algorithm that controls the quality of the reconstructed signal. 


\begin{figure}[ht]
\begin{center}
\includegraphics[width=\columnwidth]{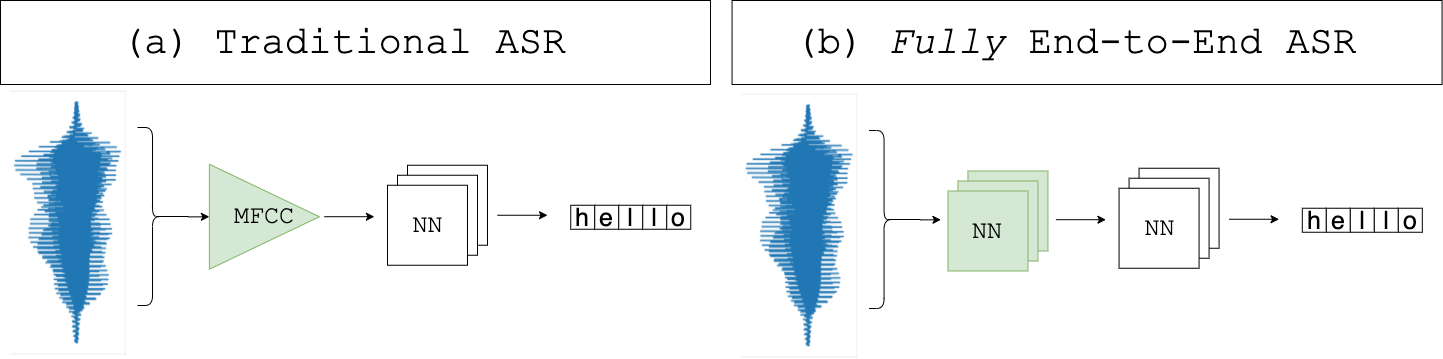}
\caption{The architectures of (a) traditional and (b) fully end-to-end ASRs. The primary difference between the two is that fully end-to-end ASRs do not include any signal processing-based feature extraction algorithms (e.g., MFCC). Instead, they use NNs that learn to extract the appropriate features during model training.}
\label{fig:asr_pipeline}
\end{center}
\end{figure}

\subsection{Feature Extraction in ASRs}

Several techniques have been used in the ASR pipeline. Here, we distinguish between {\em traditional} approaches that use signal processing algorithms for features extraction, and {\em fully end-to-end} approaches that directly learn features from data.

Traditional approaches use signal processing algorithms (e.g., STFT and
Mel-frequency Cepstral Coefficients (MFCC)~\cite{mel_scale}) to extract features from
raw audio samples (Figure~\ref{fig:asr_pipeline} (a)). The extracted features are then passed to a model for
inference, which can either be a neural network~\cite{arxiv-DS1,amodei2016deep} or a
Hidden Markov Model-Gaussian Mixture Model~\cite{kaldi_hmm,sphinx}. However,
hand-crafting features may not maximize model
accuracy~\cite{sainath2015learning}, as important features may be considered
unimportant and discarded~\cite{ilyas2019adversarial}.

In contrast, more modern approaches do not directly use traditional signal processing for feature
extraction (Figure~\ref{fig:asr_pipeline} (b)). Instead, the ASR pipeline incorporates additional (trainable) neural network layers
in the model which {\em learn} to extract features to maximize accuracy~\cite{sainath2015learning,jaitly2011learning,palaz2013estimating,tuske2014acoustic,hoshen2015speech,zeghidour2018end,zeghidour2018fully,fu2018end}. 
Thus, a {\em fully} end-to-end ASR model consists of a single element (i.e., a neural
network) that operates directly on the raw audio waveform. Since the currently published psychoacoustic attacks depend on the existence of a signal processing-based feature extraction layer, none of them work against
the newer fully end-to-end models. In this work, we propose a new psychoacoustic attack that
overcomes this limitation and produces high-quality audio against fully
end-to-end models.

\subsection{Psychoacoustic model}\label{psyc_model}
\paragraph{Audio Perception}
Psychoacoustic models have been developed to capture the functioning of the
human auditory system. In particular, psychoacoustic models tell us that there
are two phenomena that affect the perception of changes made to an audio sample.

\begin{enumerate}

    \item \textbf{Frequency Masking:}\label{frequency_masking} 
    Consider a signal that is composed of two tones. Frequency masking is when the softer tone (maskee) is inaudible in the presence of the louder tone (masker). However, increasing the tone of the maskee beyond a certain threshold can make it audible, despite the presence of a masker. As the intensity of the maskee is below the threshold, it will remain largely imperceptible.
    
   \item \textbf{Variable Perception:}\label{non_linear_precp}
    The human ear perceives audio in a variable manner. Studies in the field of human audio perception have resulted in widely familiar models of frequency perception; the Mel Scale~\cite{melscale} and the BARK Scale~\cite{barkscale}~\footnote{A visual plot of the BARK is provided in our supplementary materials.}
    , the later of which we use in our masking frequency calculations described in Algorithm~\ref{algo:attack}. Additionally, studies in this field have also produced models of perceived loudness, with the most popular being the ISO 226 Equal-Loudness Contours~\cite{iso226}.These auditory models demonstrate that lower frequencies (e.g., 100Hz to 200Hz) are perceived as less intense than higher ones (e.g., 10,000Hz to 20,000Hz) at the same actual intensity. In addition to this, it is also more difficult for humans to perceive small differences in frequency in high frequency tones as opposed to low frequency tones. As a result, lower frequencies of an audio sample can be perturbed to a relatively larger degree without resulting in perceptible distortion. In this case, distortion is an alteration that degrades the quality of the original signal.
\end{enumerate}

\begin{algorithm}[H]


\SetAlgoLined
  $\vec{x}_{\rm adv} = \vec{x}$ ; \\
  $S_{mag} = |\text{STFT}(\vec{x})|$ ; \\
  $M$ = generate\_thresholds($S_{mag}$) ; \\
  \While{$f(\vec{x}_{\rm adv}) = t$}{  
       $\vec{\delta} = \text{sign}(\nabla l(\theta, \vec{x}_{\rm adv}, t))$ \\
     $\Delta = \text{STFT}{(\vec{\delta})} $ ; \\
     \For {$i=1,2,\ldots,n$\text{ and }$j=1,2,\ldots,m$:}{
         $\Delta_{i,j} ' = |M_{i,j}|\frac{\Delta_{i,j}}{\left | \Delta_{i,j} \right |}$
      }
      
     $\vec{\delta}' = \text{Griffin-Lim}(\Delta', k)$ ; \\
     $\vec{x}_{\rm adv} = \vec{x}_{\rm adv} + \vec{\delta}'$\\

  }

 \caption{Psychoacoustic PGD Attack Steps}
 \label{algo:attack}

\end{algorithm}

\paragraph{Masking Thresholds}
Both phenomena can be accounted for through the use of~\textit{masking
thresholds}. These define the maximum degree by which each
frequency can be perturbed without any perceivable audible difference. For example, if a frequency intensity is increased beyond what is defined in the threshold, it will become audible to the human ear. Therefore, one can use this to ensure that the frequencies of the perturbations remain below the masking thresholds, thereby maintaining audio quality. 

Masking thresholds are determined using the frequencies present in an audio signal, and
thus they vary throughout the audio~\footnote{In most cases audio, specially recorded speech, the signal will consist of multiple frequencies. However, we note that it is technically possible for a recorded speech signal to consist of a singular frequency, but is highly uncommon}. We can calculate these thresholds using a
sliding window, which is moved by fixed-sized increments. This is done by
taking the magnitude of each complex coefficient of the STFT representation of a signal, which we refer to as: $[S_{\rm mag}]_{i,j} = | [S]_{i,j} | \in \mathbb{R}^{m \times n}~\refstepcounter{equation}\label{eq:smag}$

This is then followed by the calculation of the masking thresholds: 

\begin{equation*} 
    M = \text{generate\_thresholds}{(S_{\rm mag})}
\end{equation*}

Here $\vec{x}$ is an audio sample, $S_{\rm mag}$ is its magnitude spectrum and $M$ is the matrix of the corresponding masking thresholds. For more details, interested readers can refer to~\cite{lin2015principles}. 

\paragraph{Psychoacoustic Attacks}
Attacks that leverage psychoacoustic information can generate higher
quality attack audio (i.e., ones with lesser audible distortion)~\cite{schonherr2018adversarial,qin2019imperceptible}. By ensuring that the perturbation remains below the masking thresholds, the attacks produce audio whose distortions are largely imperceptible. 

\paragraph{Psychoacoustic Attacks vs Fully End-to-End ASRs}
However, a major limitation of current psychoacoustic attacks~\cite{schonherr2018adversarial,qin2019imperceptible} is that they require the presence of the traditional signal-processing-based feature extraction layer in the target ASR\footnote{We contacted the authors of current psychoacoustic attacks~\cite{schonherr2018adversarial} who confirmed our hypothesis.}. This layer could be based on either STFT or the MFCC algorithms. The attack calculates and psychoacoustically scales the gradients, which are in the frequency domain at this layer. Since psychoacoustic scaling operates in the frequency domain, the ASR pipeline must contain a traditional feature extraction layer. However, there is no such layer in an end-to-end ASR, rendering existing psychoacoustic attacks obsolete.

\section{$L_p$ Clipping and Psychoacoustics}
\label{sec:LP}
\begin{figure*}
	 \center{\includegraphics[width=\textwidth]
	      {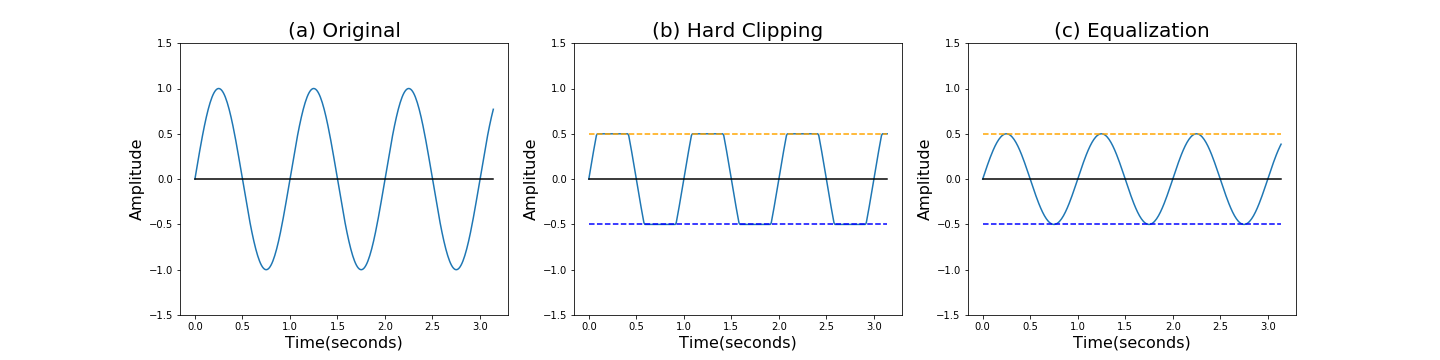}}
	  \caption{The figure above demonstrates the effects on an of (b) Hard Clipping and (c) Equalization on an (a) Original signal. We can see that (b) Hard Clipping results in a distortion of the audio signal, making the audio noisier. In contrast, (c) Equalization reduces the intensity of a single frequency in the signal, avoiding any audible distortion. It is important to note that equalization does not just reduce the intensity of the whole signal, but only a particular frequency of the signal.}
	  \label{fig:hardclip} 
\end{figure*}

 One popular method to control adversarial perturbation is the $L_p$ clipping~\cite{madry2017towards,kurakin2016adversarial, abdullahs2020sok}. It uses the \texttt{clip} to ``cut'' the amplitude of a signal beyond a maximum threshold (Figure~\ref{fig:hardclip}(b)). 
$L_p$ clipping to control the magnitude of attack perturbation~\cite{abdullahs2020sok}, which 
It has been widely
successful in the space of adversarial
images~\cite{goodfellow2016deep,madry2017towards}. However, we argue that this
method should not be applied to the audio domain by evaluating it with regards to
the principles of psychoacoustics. This will motivate an alternate method to
$L_p$ clipping.


\subsection{Hard Clipping \& Psychoacoustics}

The popular $L_p$ clipping method can be categorized as ``hard clipping'' technique. Though such techniques have seen popular use in the image space~\cite{madry2017towards,kurakin2016adversarial}, they should not be used in the adversarial audio space as they do not create imperceptible perturbations. This is due to two reasons. First, hard clipping is considered to be extreme since it cuts the signal at a maximum threshold (Figure~\ref{fig:hardclip}(b)). This process modifies the waveform of a signal, which distorts an audio signal and may introduce undesirable artifacts such as harmonics. The artifacts that are introduced by hard clipping are typically unpleasant or jarring to human listeners. 




Second, hard clipping does not account for human audio perception, thus failing to produce adversarial audio that capitalizes on the human
perception of audio. As discussed in the previous section, human audio perception is a function of psychoacoustic phenomena such as frequency masking and variable perception of loudness. Therefore, adversarial audio samples generated via hard clipping will always have greater perceivable distortion than attacks that account for psychoacoustics.



\subsection{Equalization Approach}\label{equalization}

To overcome these limitations, we propose an equalization-based method. This method will help control the adversarial audio quality, but will not introduce auditory artifacts that can degrade quality. Instead of cutting audio sample at a fixed amplitude value (Figure~\ref{fig:hardclip}(b)), equalization can change the intensity of individual frequencies in a signal (Figure~\ref{fig:hardclip}(c)). This helps avoid the undesirable audio effects introduced by the $L_p$ clipping method. If used in accordance with the psychoacoustic model, equalization can help control the adversarial perturbation so that it conforms to the human model of hearing. As a result, the perturbation can remain largely imperceptible, resulting in better quality adversarial audio. 

\subsection{Attack Details:}
Provided is a brief overview of how we compute the frequency masking threshold based on psychoacoustics for imperceptible adversarial examples. We begin the process by first identifying all of the maskers as tonal in order to ensure that the threshold that we compute can always mask out the noise. We then calculate the individual masking thresholds, which is done using frequencies with respect to the BARK scale, as the spreading functions of the masker would be similar at different ``Barks''. Finally, we compute the global masking threshold which is a combination of the individual masking thresholds and the threshold of moments of silence and pauses in speech.




\section{Adversarial Audio Attack}
We now describe an equalization-based psychoacoustic attack against ASRs. It
overcomes the limitations of the existing psychoacoustic attacks and
successfully exploits a fully end-to-end model (by ignoring the feature extraction layer). Additionally, it capitalizes on the psychoacoustics of
the human ear, thereby producing audio that is of better quality than $L_p$-clipping algorithms.

\subsection{Problem Statement and Threat Model}\label{goals}
Consider a time domain audio sample $\vec{x}$, a target transcription $y$ and an ASR model
$f(\cdot)$. The attacker's goal is to produce an adversarial audio sample $\vec{x}_{\rm adv}$ that sounds like $\vec{x}$ to the human ear, but is transcribed as $y$ by $f$. 
Specifically, the attacker is successful if the two following goals are achieved.

\begin{enumerate}
    \item \textbf{Targeted Transcription:} the ASR transcribes $\vec{x}_{\rm adv}$ as $y$ (i.e., $f(\vec{x}) \neq f(\vec{x}_{\rm adv}) = y$).
    \item \textbf{Imperceptibility:} the human ear is unable to semantically distinguish between $\vec{x}_{\rm adv}$ and $\vec{x}$.
\end{enumerate}

Following prior work~\cite{carlini2016hidden,carlini2018audio,qin2019imperceptible,schonherr2018adversarial}, we consider a white-box threat model. However, in contrast to existing psychoacoustic attacks, we do \textit{not} need the target ASR to contain a hand-crafted feature extraction layer (i.e., fully end-to-end model). 



\subsection{Attack Overview}
We craft the adversarial audio using an iterative process. Upon each iteration, we generate a perturbation $\vec{\delta}$ to a benign audio sample $\vec{x}$, resulting in $\vec{x}_{\rm adv}$:  $\vec{x}_{\rm adv} = \vec{x} + \epsilon\vec{\delta}~\refstepcounter{equation}(\theequation)\label{eq:simple_attack}$.
Here, $\epsilon$ is a scalar between 0 and 1 that controls the magnitude of the perturbation. The perturbation $\vec{\delta}$ shares the dimensions of the audio $\vec{x}$ and is generated using the following two steps: 
\begin{enumerate}
    \item \textbf{Generating a candidate perturbation:} 
    Using the gradient of the model loss (on the target transcription $y$) with respect to the input audio, we produce a \textit{candidate} perturbation $\vec{\delta}$. Adding this perturbation to the current version of the audio results in an output closer to the target transcription in the decision space. 
    
    \item \textbf{Constraining the perturbation:} We
constrain the candidate perturbation $\vec{\delta}$ according to the masking thresholds to ensure that the perturbation is largely imperceptible. 
\end{enumerate}


\subsection{Attack Formulation}

For the first step, we generate a candidate perturbation by looking at the sign of the gradient of the loss with respect to the input audio. For the second step, we propose a novel set of equalization-based
techniques to constrain the perturbation according to psychoacoustics. 


\paragraph{Generating Perturbation}

To produce a candidate perturbation $\vec{\delta}$, we use the Fast Gradient Sign Method (FGSM)~\cite{goodfellow2014explaining}: 
 $\vec{\delta} = \text{sign}(\nabla l(\theta, \vec{x}, y))~\refstepcounter{equation}$.
Here, $\theta$ denotes the model's parameters, $l(\cdot)$ is the loss function, and $\nabla$ denotes the gradient so that $\nabla l(\theta, \vec{x}, y)$ is the gradient of the loss with respect to the input. 




\paragraph{Constraining Perturbation}

As done in prior work~\cite{abdullahs2020sok}, a candidate perturbation $\vec{\delta}$
could simply be added to the audio sample, scaled by some parameter $\epsilon$
(Equation~\ref{eq:simple_attack}). This scalar multiplication results in equal
scaling of all frequencies in the perturbation signal by $\epsilon$. However, this would
ignore psychoacoustics as the human ear perceives frequencies variably, resulting in a noisy perturbation. To better reflect psychoacoustics, lower
frequencies should be scaled with a higher constant since lower frequencies are
perceived as softer than higher ones. This can be achieved by first constraining the frequencies of the candidate perturbation $\vec{\delta}$ according to the masking
thresholds of the psychoacoustic model. This is followed by the application of Equation~\ref{eq:simple_attack}. This will help achieve the desired quality of audio:


\begin{enumerate}

    \item \textbf{Frequency Representation:} We constrain the perturbation $\vec{\delta}$ using the masking thresholds $M$, which we precompute according to the psychoacoustic model. The thresholds will define the maximum allowable intensity by which we can perturb each frequency in the audio without having an impact on the audio quality. 
    However, we cannot directly use the masking threshold for two reasons. The masking thresholds are in the (magnitude-only) time-frequency domain (due to the application of $S_{\rm mag}$ shown in Equation~\ref{eq:smag}), whereas the perturbation $\vec{\delta}$ is in the time domain (Equation~\ref{eq:simple_attack}). 
    
    We apply the STFT to transform $\vec{\delta}$ to the time-frequency domain. The STFT
maintains the phase information, which is an integral part of accurately
reconstructing an audio sample from its frequency representation. Reconstructing
a time-domain audio sample using only the magnitude information can result in
distortion. To produce adversarial audio that best adheres to the principals of
psychoacoustics, the phase information must be maintained while making
modifications to the spectrogram representation of an audio sample. Therefore, we use the $S$ instead of $S_{\rm mag}$: $\Delta = S(\vec{\delta}) \in \mathbb{C}^{m \times n}~\refstepcounter{equation}$.
Here, $\Delta$ is the spectrogram for perturbation $\vec{\delta}$. 


    \item \textbf{Equalization:} Now, we use equalization (\ref{equalization}) to constrain the perturbation according to the masking thresholds $M$ in the frequency domain. We equalize the spectrogram of the perturbations ($\Delta$) for the indices $i,j$ that violate the masking thresholds. 
    We scale the real-imaginary pair of the complex number, thereby maintaining the phase:
       \begin{equation} \label{equlaization}
     \Delta_{i,j} ' = |M_{i,j}|\frac{\Delta_{i,j}}{\left | \Delta_{i,j} \right |} 
     \in \mathbb{C}^{m \times n}
    \end{equation}  
    where $\Delta_{i,j} '$ is the scaled value, $i$ the frequency bin, and $j$ is the time index. Since Equation~\ref{equlaization} is equalizing using the masking thresholds $M$, this method will produce higher quality audio, while avoiding the distortion that is a result of $L_p$ based clipping methods. Additionally, this method will work in the absence of a traditional feature extraction layer, thereby overcoming the limitation of existing psychoacoustic attacks.

    \item \textbf{Time Domain Reconstruction:} Since $\Delta'$ is a \textit{modified} time-frequency representation of $\Delta$, it does not have a perfect time-domain representation (Section~\ref{stft}). We obtain an approximate reconstruction via the Griffin-Lim algorithm (Section~\ref{stft}). The reconstructed time-domain perturbation $\vec{\delta}'$ is then added to the original audio sample to produce an adversarial sample.

\end{enumerate}






\begin{table*}


\centering
\resizebox{\textwidth}{!}{%
\begin{tabular}{@{}l|ccc|cccc@{}}
\toprule
\multicolumn{1}{c|}{} &
  \multicolumn{3}{c|}{{
  \textit{\textbf{Which one is noisier?}}}} &
  \multicolumn{4}{c}{{
  \textit{\textbf{How does the noisy audio sample (CW) differ from the other one (us)?}}}} \\ \cmidrule(l){2-8} 
\multicolumn{1}{c|}{\multirow{-2}{*}{\textit{\textbf{Audio Samples}}}} &
  {
  \textit{\textbf{\begin{tabular}[c]{@{}c@{}}Noisy CW\\ (\%)\end{tabular}}}} &
  {
  \textit{\textbf{\begin{tabular}[c]{@{}c@{}}Noisy Us\\ (\%)\end{tabular}}}} &
  {
  \textit{\textbf{\begin{tabular}[c]{@{}c@{}}Both sound same\\ (\%)\end{tabular}}}} &
  {
  \textit{\textbf{\begin{tabular}[c]{@{}c@{}}Perceptible but not annoying \\ (\%)\end{tabular}}}} &
  {
  \textit{\textbf{\begin{tabular}[c]{@{}c@{}}Slightly annoying \\ (\%)\end{tabular}}}} &
  {
  \textit{\textbf{\begin{tabular}[c]{@{}c@{}}Annoying \\ (\%)\end{tabular}}}} &
  {
  \textit{\textbf{\begin{tabular}[c]{@{}c@{}}Very annoying \\ (\%)\end{tabular}}}} \\ \midrule
That is comparatively nothing &
  83 &
  10 &
  7 &
  30.12 &
  48.19 &
  20.48 &
  1.20 \\
Talking later is beneath us &
  89 &
  6 &
  5 &
  5.62 &
  29.21 &
  44.94 &
  20.22 \\
But there seemed no &
  97 &
  3 &
  0 &
  10.31 &
  36.08 &
  41.24 &
  12.37 \\
Been looking up tombs county &
  83 &
  3 &
  14 &
  32.53 &
  50.60 &
  16.87 &
  0 \\
Foul mouth fellow at the top &
  93 &
  6 &
  1 &
  2.15 &
  27.96 &
  37.63 &
  32.26 \\
Tied to a woman &
  85 &
  6 &
  9 &
  27.06 &
  49.41 &
  17.65 &
  5.88 \\ \bottomrule
\end{tabular}%
}
\caption{Breakdown of participants' selection of noisier audio across the audio
samples and subjective assessment of noise difference in a pair of audio
samples. More than 80\% of participants selected each of the CW's audio
samples to be noisier than ours (left). Participants' subjective audio
perception 
rating between CW and our audio samples are recorded(right). About 70\% - 98\% (aggregating ratings for slightly annoying, annoying, and very annoying) of the participants found CW's audio samples to be annoying.}
\label{tab:audio-comparison}
\vspace{-2em}
\end{table*}

\section{Experimental Setup}

\subsection{ASR Model}
To demonstrate that our attack exhibits cross-architecture generalization, we run it against both a traditional model and a fully end-to-end one. For the traditional ASR, we use DeepSpeech\footnote{Code available at: \texttt{https://github.com/mozilla/DeepSpeech}} and for the fully end-to-end one, we employ Wav2Letter~\cite{collobert2016wav2letter}.

\subsubsection{DeepSpeech} 
The model pipeline consists of an STFT stage followed by convolutions and bi-directional RNNs. The model was trained on the LibriSpeech data set to achieve the state-of-the-art word error rate of 8\%~\cite{mozilla_ds_0.4.1}.

DeepSpeech is not a fully end-to-end model because it includes a signal processing-based feature extraction layer (i.e., STFT). However, we can treat it as such by simply ignoring the STFT stage (as if it does not exist) and applying the attack directly on the raw audio. This highlights that our attack works against both types of models (i.e., those that are fully end-to-end and those that are not). In other words, our attack {\em does not require the signal processing based feature extraction layer}.

\subsubsection{Wav2Letter} 
This is a fully end-to-end model consisting of stacked layers of convolutions (without any recurrent layer)\footnote{Despite the lack of a recurrent layer, which is commonly found in most ASR architectures, we specifically chose Wav2Letter due to an explicit request by a previous reviewer. This is important to note, since the lack of the RNN might reduce attack success for optimization attacks (Section~\ref{succ})}. We trained this model on the LibriSpeech data set to achieve the state-of-the-art word error rate of 7\%. Since there is no signal processing-based feature extraction layer, this model is not vulnerable to existing psychoacoustic attacks~\cite{qin2019imperceptible, schonherr2018adversarial}.

\subsection{Dataset and Attack Parameters}
Following the experimental setup outlined in Qin et al~\cite{qin2019imperceptible}, we randomly sample 100 audio samples from the LibriSpeech~\cite{panayotov2015librispeech} test set, each of which is perturbed through our attack using 1000 iterations. Each audio sample is perturbed to produce 50 random target phrases sampled from the test set. The attack is successful if the ASR transcription \textit{exactly} matches the target transcription.

We also repeat this experiment for different values of $k$, which is the number of iterations of the Griffin-Lim algorithm\footnote{Code available at: \texttt{https://librosa.github.io/ librosa/generated/librosa.core.istft.html.}}. This parameter influences the quality of audio reconstruction and thus has an impact on perceptibility and attack success. Specifically, we repeat the above-described attack experiment for $k=1,2,4,8$. This results in a total of 20,000 attack audio samples (100 audio samples $\times$ 50 target transcriptions $\times$ 4 iteration values) for the entire experimental setup. Such a large number of attack audio files were generated to thoroughly test the viability of our attack.


\subsection{User Study}
\label{user-study}

We ran a user study to determine whether our attack perturbation are less noisy
than other attacks by comparing it to that of state-of-the-art CW~\cite{carlini2018audio}. We chose this attack since it is the only other one that does work against the fully end-to-end model. However, unlike our attack, CW's attack ignores psychoacoustics when perturbing adversarial audio samples. 

We hypothesized that participants who listened to both audio samples would find
our audio to be less noisy. To test this, 100 participants were recruited using
Amazon's Mechanical Turk (MTurk) and compensated two USD upon completion of a Qualtrics survey. They had to read and digitally sign a consent form before they could
continue. Then, participants were asked to disclose any
hearing issues; however, this did not disqualify them from our study. 

The survey contained ten questions
about audio samples from our and CW's attack~\cite{carlini2018audio}. Participants were presented with two audio samples (Audio Sample A 
and Audio Sample B) and asked which of the two samples was
noisier. Participants were given the option to choose (A) Audio Sample A; (B) Audio
Sample B; or (C) The two pieces of audio are the same. If users selected A or B,
they were asked a follow-up question: ``How does the noisy audio differ from the
other one?'' Options were (A) Perceptible but not annoying; (B) Slightly
Annoying; (C) Annoying: (D) Very Annoying. 

In four of the ten questions, Audio Sample A and Audio Sample B were identical, where two were our audio, and two were Carlini's. This was done to
test the attentiveness of participants while they were answering the questions. The remaining six
questions compared CW's audio to our audio. In these cases the
transcription of the audio was the same; only the technique to produce the audio was
different.  Finally, we asked participants to disclose the device (e.g., laptop, desktop, or other) and headphones types. 

\section{Experimental Results}
To demonstrate the viability of our attack against ASRs, we needed to
demonstrate that the two previously outlined attacker goals are met (Section~\ref{goals}).
Specifically, that the rate of successful target transcription and the imperceptibility of
the perturbations used in our attack are either equal to or better than other attacks in current literature.

\subsection{Attack Success Rate}\label{succ}

For all the attack audio files generated with $k=1$ and $2$, the ASR
transcribed them as malicious text. When attacking DeepSpeech, our attack was able to achieve a success rate of 100\%, which is on par with other attacks in this space~\cite{carlini2016hidden,carlini2018audio,schonherr2018adversarial}. However, the success rate with $k=1$ against Wav2Letter model fell to 76\%. Upon further investigation, we observed that the success rate for CW also failed to reach 100\%. We believe that this was due to the lack of a recurrent layer or the MFCC in the Wav2Letter architecture, which lead to reduced attack success. Additionally, we discuss why smaller values of $k$ had a higher success rate than larger ones in the supplementary material.

The number of iterations of the Griffin-Lim algorithm influences both the rate
of attack success and the time it takes to generate the audio file. We discuss
in the supplementary materials how we converge to the optimum value for this variable
via empirical tests and manual listening experiments. We found this value to be
$k=1$, and will be using it for the remainder of the experiments.

These results demonstrate the steps of our psychoacoustic attack still converge to the desired solution, with the same accuracy as that of other attacks. This is true even though our attack requires switching between domains (time-frequency) and involves approximating a modified STFT. The time-frequency conversion of the STFT is lossy and equalization is happening in the frequency domain. As a result, if the attack steps are structured the way they are, they would have introduced additional loss and have had prevented convergence.

\subsection{User Study}
In our MTurk study, 100 participants gave consent and completed the survey entirely. Among them, 61 were male, and 39 were female. The average age of
the participants was 35.5. All participants
reported being native English speakers. Each of the participants
used either a laptop or a desktop for the study and wore a headset for the audio experiment. The median completion time of the study was
3.62 minutes. During the study, each participant
was presented with ten pairs of audio samples in random order. If not stated otherwise, the rest of this
section presents results made from observations from comparisons of the six pairs of audio samples, one from CW's attack and one from ours. (For details,
please see Section~\ref{user-study}.)

Table~\ref{tab:audio-comparison} shows the results of our study. More than 80\% of
participants considered \underline{all} of CW's attack audio samples as noisy. Note that across all samples, the number of participants to consider the CW audio noisier was at least 83\%. Table~\ref{tab:audio-comparison} also
shows the breakdown of audio perception difference for participants who selected
CW to be noisier among the two audio samples. We can see that about 70\% -
98\% (aggregating ratings for slightly annoying, annoying and very annoying) of
the participants found CW's audio samples to be annoying. Meanwhile, only a
small proportion (2\% - 32\%) perceived the audio sample generated by CW's
technique to be distinguishable from ours but they did not find it annoying. 

To measure the statistical significance of our observations, we ran a $\chi^2$
with the  null hypothesis: there is no audio quality difference between
CW's and our attack audio samples. With a
\textit{p-value} of less than 0.01, the null hypothesis is rejected and thus shows that there exists a significant noise quality difference between CW and our audio samples.




\section{Conclusion}

ASRs are vulnerable to adversarial examples. The more
potent attacks in this space are the targeted ones based on psychoacoustics.
These can produce clean audio samples (i.e., largely imperceptible perturbation)
that are transcribed to a targeted transcription. Though these attacks are
effective against traditional ASRs, they are obsolete against newer, fully
end-to-end ones. This is due to the absence of the traditional,
signal-processing-based feature extraction layer.
In this paper, we propose an equalization-based attack that
leverages signal processing and psychoacoustics to produce
clean adversarial audio. As a result, our attack produces less noisy audio than the current state-of-the-art attacks. In the process of developing our attack, we discovered that the $L_p$ clipping method, which has been used in the past~\cite{abdullahs2020sok}, is a poor technique for controlling the imperceptibility of a perturbation. Our work has a 100\% success rate, produces high-quality audio, and is applicable to both traditional and fully end-to-end ASRs. 

\section{Acknowledgments}
This work is partly supported by the National Science Foundation under CNS-1933208.
Any opinions, findings, and conclusions or recommendations
expressed in this material are those of the authors and do
not necessarily reflect the views of the National Science Foundation.
\bibliography{refs,defs,misc,speech_rec}
\appendix
\section{Supplementary Material}



      




\subsection{Attack Parameters}\label{gl_calc}
\begin{figure}%
    \centering
    \subfigure[Attack Success vs Griffin-Lim Iterations]{{\includegraphics[width=.4\linewidth]{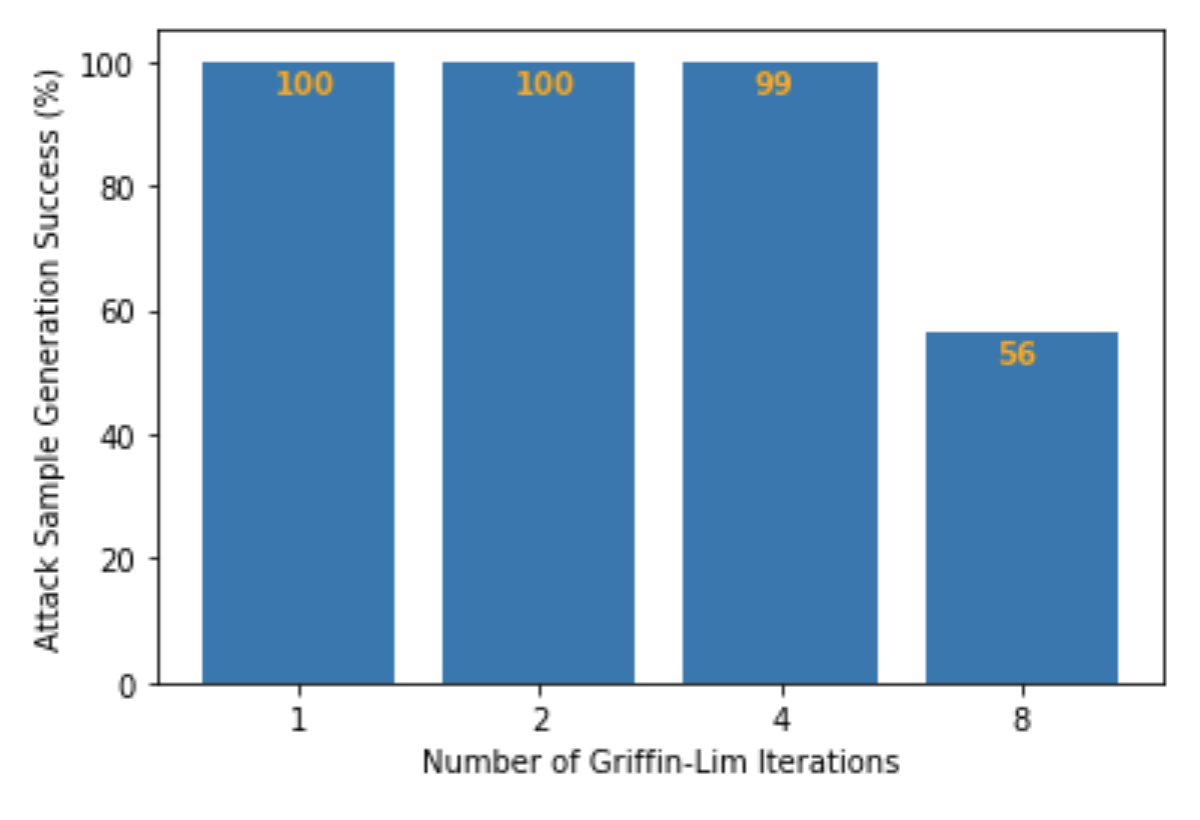}}}%
    \qquad
    \subfigure[Attack Iterations  vs Griffin-Lim Iterations ]{{\includegraphics[width=0.4\linewidth]{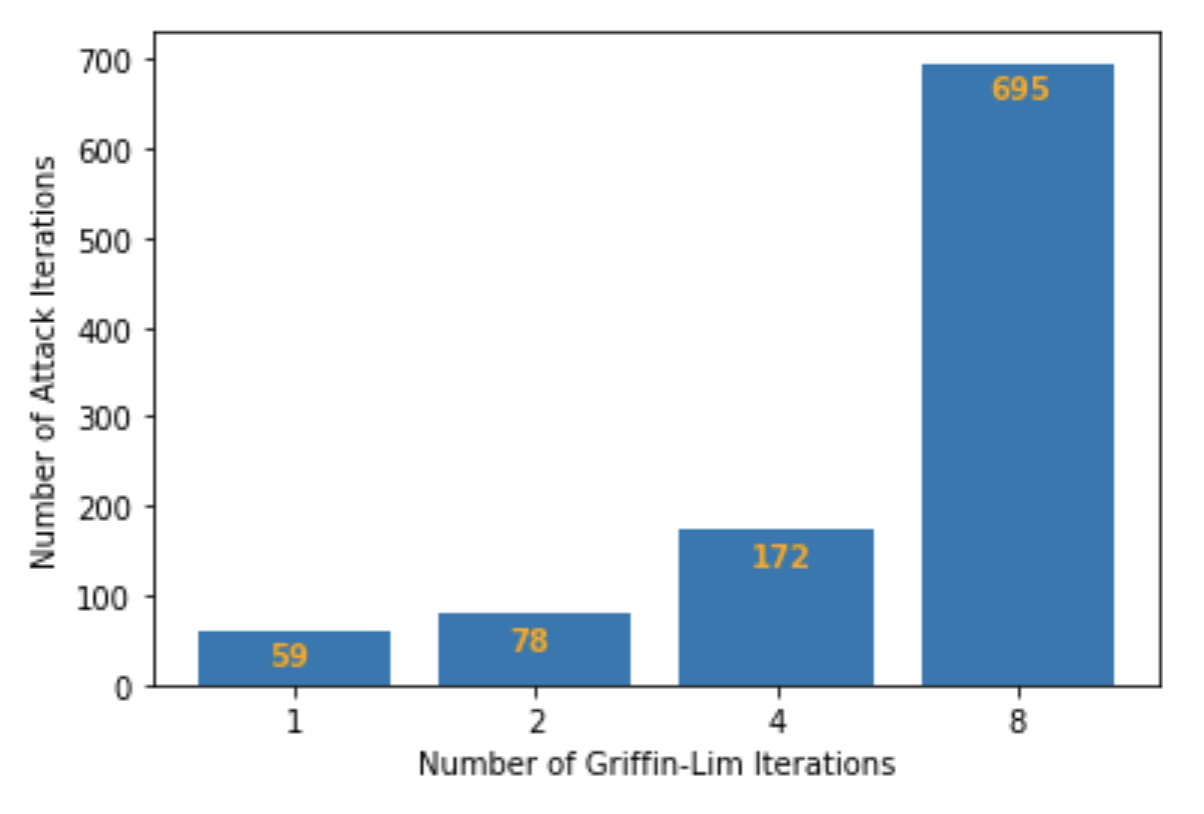} }}%
    \caption{ The plot shows the iterations of the Griffin-Lim algorithm that would be ideal for an attacker. (a) Increasing the Griffin-Lim iterations leads to a reduction in attack success. (b) Higher Griffin-Lim iterations requires a higher number of attack iterations to produce an attack audio sample. Manual listening tests showed that audio quality remained consistent across all Griffin-Lim iterations.}%
    \label{fig:results}%
\end{figure}

The attack success decreases if we increase the number of iterations beyond 2 (Figure~\ref{fig:results}(a)). In other use cases, the Griffin-Lim algorithm is executed for 50 iterations~\cite{wang2017tacotron}. However, these number of iterations are too high for our use case, as we see a significant drop in accuracy (approximately 50\%) at just 8 iterations (Figure~\ref{fig:results}(a)). Similarly, increasing the number of Griffin-Lim iterations also results in an increases the number of required iterations of our attack (Figure~\ref{fig:results}(b)), hence a longer required time to generate an attack audio sample. Clearly, increasing the number of Griffin-Lim iterations is unhelpful for the attacker. As a result, it is worth discussing why this is the case.

As discussed in Section 3.1, the Griffin-Lim algorithm provides an approximate reconstruction of time-domain audio, since a perfect reconstruction is not possible. Increasing the number of iterations results in the finer perturbations being discarded. This decreases the likelihood of the adversarial audio moving in the direction of the target transcription in the decision space. 

Moreover, we found that increasing the number of Griffin-Lim iterations did not actually improve the quality of the adversarial audio samples. Manual listening tests by the authors revealed that the perceived quality of adversarial audio samples was consistent across the number of Griffin-Lim iterations. As a result, an attacker will benefit from setting the number of Griffin-Lim iterations to 1. This will allow for a high success rate of adversarial audio with fewer attack iterations, hence faster attack audio generation. Specifically, increasing the Griffin Lim algorithm from 1 to 8 will increase the total attack iterations (and consequently the compute time) by 11 times. 

\begin{figure}
      \centering
    \includegraphics[width=8cm]{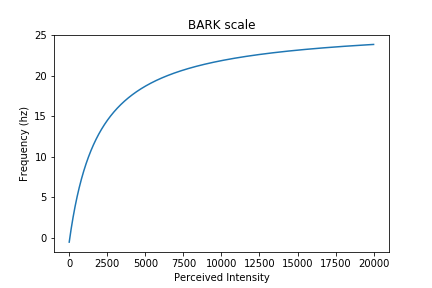}
    \caption{The BARK scale is a psycoacoustic model of the human perception of loudness in relation to frequency. The scale suggests that humans perceive high frequencies as louder than lower frequencies. The perceived intensity values associated with each frequency correspond to the first 24 critical bands of hearing.}
    \label{fig:BARK_scale} 
\end{figure}

\end{document}